%% file: main.tex
\documentclass[acmsmall]{acmart}

\AtBeginDocument{%
  \providecommand\BibTeX{{%
    Bib\TeX}}}

\setcopyright{cc}
\setcctype{by}
\acmDOI{10.1145/3797126}
\acmYear{2026}
\acmJournal{PACMSE}
\acmVolume{3}
\acmNumber{FSE}
\acmArticle{FSE031}
\acmMonth{7}
\acmSubmissionID{fse26maina-p637-p}
\received{2025-09-12}
\received[accepted]{2025-12-22}

\DeclareMathDelimiter{(}{\mathopen} {operators}{"28}{largesymbols}{"00}
\DeclareMathDelimiter{)}{\mathclose}{operators}{"29}{largesymbols}{"01}

\usepackage{booktabs}
\usepackage{multirow}
\usepackage{subcaption}
\usepackage{graphicx}
\usepackage{graphbox}
\usepackage{setspace}
\usepackage{wrapfig}
\usepackage{floatflt}
\usepackage{xspace}

\usepackage[colorinlistoftodos,prependcaption,textsize=tiny]{todonotes}
\usepackage{verbatim}

\usepackage{graphicx}
\usepackage{color}
\usepackage{url}
\usepackage{listings}
\usepackage{csquotes}
\usepackage[normalem]{ulem}
\usepackage{setspace}
\usepackage{algorithm}
\usepackage[most]{tcolorbox}
\usepackage[noend]{algpseudocode}

\usepackage{boldline}
\usepackage{cellspace}
\usepackage{ragged2e}
\usepackage{colortbl}

\algblock{Input}{EndInput}
\algblock{Output}{EndOutput}

\usepackage{xparse}
\usepackage{mathtools}
\usepackage{multicol}
\usepackage{adjustbox}
\usepackage{threeparttable} 
\usepackage{enumitem}
\def\BibTeX{{\rm B\kern-.05em{\sc i\kern-.025em b}\kern-.08emT\kern-.1667em\lower.7ex\hbox{E}\kern-.125emX}}

\usepackage{textcomp}
\usepackage{hyperref}
\usepackage{float}

\newcommand{\mybox}[1]{\begin{tcolorbox}[enhanced, frame hidden, boxsep=0pt]\emph{#1}\end{tcolorbox}}

\newcommand{\remove}[1]{}
\newcommand{\git}[1]{\textit{GitHub}}

\newcommand{\cmr}[1]{#1}

\usepackage{eqparbox}
\newdimen{\algindent}
\algnewcommand\LeftComment[2]{%
\hspace{#1\algindent}$\triangleright$ #2 \hfill %
}

\usepackage{xcolor}

\definecolor{codegreen}{rgb}{0,0.6,0}
\definecolor{codegray}{rgb}{0.5,0.5,0.5}
\definecolor{uncoveredgray}{rgb}{0.8,0.8,0.8}
\definecolor{codepurple}{rgb}{0.58,0,0.82}
\definecolor{backcolour}{rgb}{0.99,0.80,0.87}

\newcommand{\parabf}[1]{\noindent\textbf{#1}}

\lstdefinestyle{demo-code}{
    backgroundcolor=\color{white},   
    commentstyle=\color{codegreen},
    keywordstyle=\color{magenta}, 
    numberstyle=\tiny\color{codegray},
    stringstyle=\color{codepurple},
    breakatwhitespace=false,         
    breaklines=true,                 
    captionpos=b,                    
    keepspaces=true,                 
    numbersep=5pt,                   
    showspaces=false,                
    showstringspaces=false,
    showtabs=false,                  
    tabsize=2,
    xleftmargin=0.05\columnwidth,
    xrightmargin=0.05\columnwidth,
}

\newcommand{\swebench}{SWE-bench}

\newcommand{\agentless}{\textsc{Agentless}}
\newcommand{\sweagent}{SWE-agent}
\newcommand{\sweagentmini}{mini-SWE-agent}
 
\newcommand{\tools}{\textsc{Claude-Tools}} 
\newcommand{\acr}{\textsc{AutoCodeRover}} 
\newcommand{\ssi}{\textsc{SWESynInfer}} 
\newcommand{\ct}{\textsc{Claude Tools}} 
\newcommand{\flt}{\textit{Frame Lifetime Trace}}

\newcommand{\claudesonnet}{Claude-Sonnet-3.7}
\newcommand{\gpto}{GPT-4o}

\newcommand{\qwen}{Qwen3}

\newcommand{\claudethreefive}{Claude-Sonnet-3.5}

\newcommand{\baseagent}{BaseAgent}
\newcommand{\pdbagent}{BaseAgent$_{pdb}$}
\newcommand{\adi}{Agent-centric Debugging Interface}
\newcommand{\adis}{ADI}

\newcommand{\dbgagent}{\textsc{FramePilot}\xspace}

\usepackage{caption}

\begin{document}

\title{Empowering Autonomous Debugging Agents with Efficient Dynamic Analysis}

\author{Jiahong Xiang\textsuperscript{\textdagger{}}}
\orcid{0009-0008-5239-2540}
\affiliation{
    \institution{Research Institute of Trustworthy Autonomous Systems, Southern University of Science and Technology}
  \city{Shenzhen}
  \country{China}}
\email{xiangjh2022@mail.sustech.edu.cn}

\author{Xiaoyang Xu}
\orcid{0009-0006-5707-6665}
\affiliation{%
  \institution{Southern University of Science and Technology}
  \city{Shenzhen}
  \country{China}
}
\email{12112620@mail.sustech.edu.cn}

\author{Xiaopan Chu}
\orcid{0009-0005-8573-3549}
\affiliation{%
  \institution{Southern University of Science and Technology}
  \city{Shenzhen}
  \country{China}
}
\email{12111712@mail.sustech.edu.cn}

\author{Hongliang Tian}
\orcid{0009-0005-1248-4078}
\affiliation{%
  \institution{Ant Group}
  \city{Hangzhou}
  \country{China}
}
\email{tate.thl@antgroup.com}

\author{Yuqun Zhang\textsuperscript{\textdagger{}}*}
\orcid{0000-0003-2239-6723}
\affiliation{%
  \institution{Research Institute of Trustworthy Autonomous Systems, Southern University of Science and Technology}
  \city{Shenzhen}
  \country{China}}
\email{zhangyq@sustech.edu.cn}

\thanks{\textsuperscript{*}Yuqun Zhang is the corresponding author.}
\thanks{\textsuperscript{\textdagger{}}These authors are also affiliated with the Department of Computer Science and Engineering, Southern University of Science and Technology, Shenzhen, China.}

\begin{abstract}
Autonomous agents for automated program repair represent a promising frontier in software engineering, yet their effectiveness is often hindered by reliance on post-mortem, coarse-grained execution feedback. While integrating traditional interactive debuggers seems a natural solution, their low-level, line-by-line interaction paradigm turns \cmr{out} to be cost-inefficient for LLM-based agents, leading to exhausted budgets and unproductive loops. To mitigate this, we introduce \adi{} (\adis), a novel agent-centric debugging interface designed for cost-efficient, end-to-end autonomous interaction. Specifically, \adi{} realizes a function-level interaction paradigm, powered by our \flt{}—a comprehensive data structure encapsulating a function's stateful execution trace—and a set of high-level navigational commands.

Our extensive evaluation on the \swebench{} benchmark demonstrates the effectiveness and efficiency of \adis{}. By simply equipping a basic agent with \adis{}, it successfully resolves 63.8\% of the tasks on the \swebench{}-Verified set, even slightly outperforming the highly-optimized and high-investment \tools{} agent, at an average cost of \$1.28 per task with \claudesonnet{}. 
Furthermore, we demonstrate \adis{}'s generality by integrating it as a plug-and-play component into the existing SOTA agents, delivering consistent gains ranging from 6.2\% to 18.5\% on the resolved tasks. These results indicate that \adi{} could achieve a general and efficient enhancement for the existing autonomous agents.

\end{abstract}




\maketitle

\input{introduction}

\input{overview}

\input{background}

\input{approach}

\input{evaluation}

\input{threats}

\input{conclusion}

\newpage
\clearpage
\bibliographystyle{ACM-Reference-Format}
\bibliography{dbgref}
\end{document}

%% file: introduction.tex

\section{Introduction}

Recently, Large Language Models (LLMs) have demonstrated remarkable capabilities in addressing a wide range of software engineering challenges~\cite{xia2023automated,codamosa,swebench,specrover,libro,titanfuzz,li2023can,hou2024large, tan2024llm4decompile, tan2025decompile, tan2025sk2decompile, tan2024prompt, xiang2026evaluating, xiang2024far, tan2024hicl, zeng2022extensive, he2026reflair}. A particularly promising frontier is the development of autonomous agents for automated program repair ~\cite{, sweagent, specrover, openhands, liu2024large, li2025patchpilot, bouzenia2024repairagent, barke2023grounded, tang2024study, chatrepair}, which are increasingly benchmarked against complex, repository-level real-world tasks like those in the \swebench{} suite~\cite{swebench}. 
Specifically, these automated program repair tasks are highly demanding, requiring an agent to possess capabilities like \cmr{issue reproduction}, fault localization, and patch generation. 
In particular, the existing state-of-the-art agents primarily employ an iterative debugging process through self-correction based on execution feedback~\cite{sweagent, openhands, specrover, selfdebug, chatrepair}. However, relying solely on post-mortem~\cite{WilkesWheelerGill1951}, coarse-grained execution output for self-correction causes limited effectiveness for fixing complex bugs, which are typically involved in intricate data and control flow relationships that are often not presented in the final execution results. For instance, a simple failure signal (e.g., an assertion error) or a non-crashing unexpected behavior (e.g., an incorrect output) caused by a complex bug often reveals little information about the state-dependent execution path leading to the failure, i.e., providing insufficient assistance for LLMs to locate the bug and further generate the patch accordingly, thus compromising the power of applying LLMs~\cite{debuggym, chatdbg, autosd}. 

To mitigate the above-mentioned issue, one can be inspired by the existing human-in-the-loop techniques, i.e., using interactive debuggers such as GDB~\cite{gdb}, PDB~\cite{pdb}, or those integrated into IDEs, to inspect a program's state and understand its complex data and control flow. Existing research attempts to integrate agents with these traditional, human-in-the-loop tools for both enabling fine-grained dynamic analysis and reducing the need for manual intervention, e.g., building interfaces where an agent assists human developers by answering high-level queries through autonomous control of a debugger~\cite{chatdbg} and developing interactive, text-based environments where, guided by a \cmr{developer-written} set of test cases, an agent is provided with debugger access to train and evaluate its ability to autonomously repair code~\cite{debuggym}. 
Ideally, an agent could leverage these debuggers to efficiently resolve complex debugging tasks with the assistance from the underlying human developer skills. However, while it is natural that the existing human-in-the-loop debuggers operate on a line-by-line basis using low-level, atomic commands (e.g., \texttt{next} and \texttt{print var}), such a paradigm renders the usage of an LLM agent rather cost-inefficient, thus making the agent-debugger integration somewhat ineffective in the real world. Specifically, each atomic command provides only a sliver of state information but incurs the  substantial cost of a complete LLM inference cycle. When attempting to operate fully autonomously, agents easily exhaust their computational budget or halt in unproductive debugging loops, preventing them from reaching a solution~\cite{debuggym, enigma}. As a result, existing techniques can hardly be fully autonomous, i.e., they have to rely on workarounds such as human-in-the-loop guidance~\cite{chatdbg} or operate under strong assumptions such as high-quality assertions and test cases~\cite{debuggym, autosd, vuldebugger}, posing a pressing need for agent-centric interactive debugging interfaces.

In this paper, we introduce \adi{} (\adis{}) to realize cost-efficient end-to-end autonomous agent-debugger interaction, freeing agents from reliance on human-in-the-loop guidance. Our approach is established on two key designs. First, instead of the traditional statement-level agent-debugger interaction, \adis{} realizes the function-level agent-debugger interaction via the \textit{\flt{}} (FLT), a comprehensive data structure that encapsulates the stateful execution trace of a single function invocation. Specifically, an FLT contains the function's arguments, its final return value, a complete, ordered trace of all executed statements along with their resulting state modifications, and identifiers linking to its caller and any downstream callees. 
Second, \adis{} equips an agent with a set of high-level navigational commands designed for efficient exploration of a program's dynamic states. These commands empower the agent to strategically control execution flows by setting function-level conditional breakpoints and gain a high-level overview of program structures by navigating the associated dynamic call graphs.

We perform an extensive study to evaluate the effectiveness and efficiency of \adis{}. Our evaluation shows that by simply equipping a basic agent with \adis{}, the resulting agent, \dbgagent{}, achieves \cmr{superior} performance. Specifically, it resolves 63.8\% of tasks on the \swebench{} Verified benchmark~\cite{OpenAI2024SWEBench}, even slightly outperforming the highly-optimized and high-investment \tools{} agent, at an average cost of \$1.28 per task with \claudesonnet{}~\cite{anthropic2025claude3_7}. 
To demonstrate the generality of \adis{}, we integrate it as a plug-and-play component into two SOTA agents with distinct architectures, \sweagentmini{}~\cite{swebench} and \acr{}~\cite{specrover}. This integration delivers consistent performance gains ranging from 6.2\% to 18.5\% in terms of the resolved tasks at a modest cost, showcasing \adis{}'s utility as a general-purpose enhancement. 

In summary, this paper makes the following contributions:

\begin{itemize}[leftmargin=*]
\item \textbf{\adi{}.} We propose \adi{}, a novel interactive debugging interface designed for autonomous agents that enables cost-efficient dynamic analysis through a function-level interaction model, powered by our \flt{} and a set of high-level navigational commands.

\item \textbf{\cmr{Superior} Performance.} Our extensive evaluation on the \swebench{} benchmark shows that by simply equipping a basic agent with \adi{}, it achieves \cmr{superior} performance, resolving 63.8\% of tasks in \swebench{} Verified, even slightly outperforming the highly-optimized and high-investment \tools{} agent, at an average cost of \$1.28 per task with \claudesonnet{}. 

\item \textbf{General-Purpose Enhancement.} We demonstrate the generality of \adis{} by integrating it as a plug-and-play component into two SOTA agents with distinct architectures, achieving consistent performance gains from 6.2\% to 18.5\% at a modest cost. This indicates that \adis{} could potentially enhance the performance of general agent architectures for automated program repair tasks. 

\end{itemize}

%% file: overview.tex
\section{Motivating Example}
\label{sec:motivation}

\begin{figure}[htb]
    \centering
    \includegraphics[width=0.95\columnwidth]{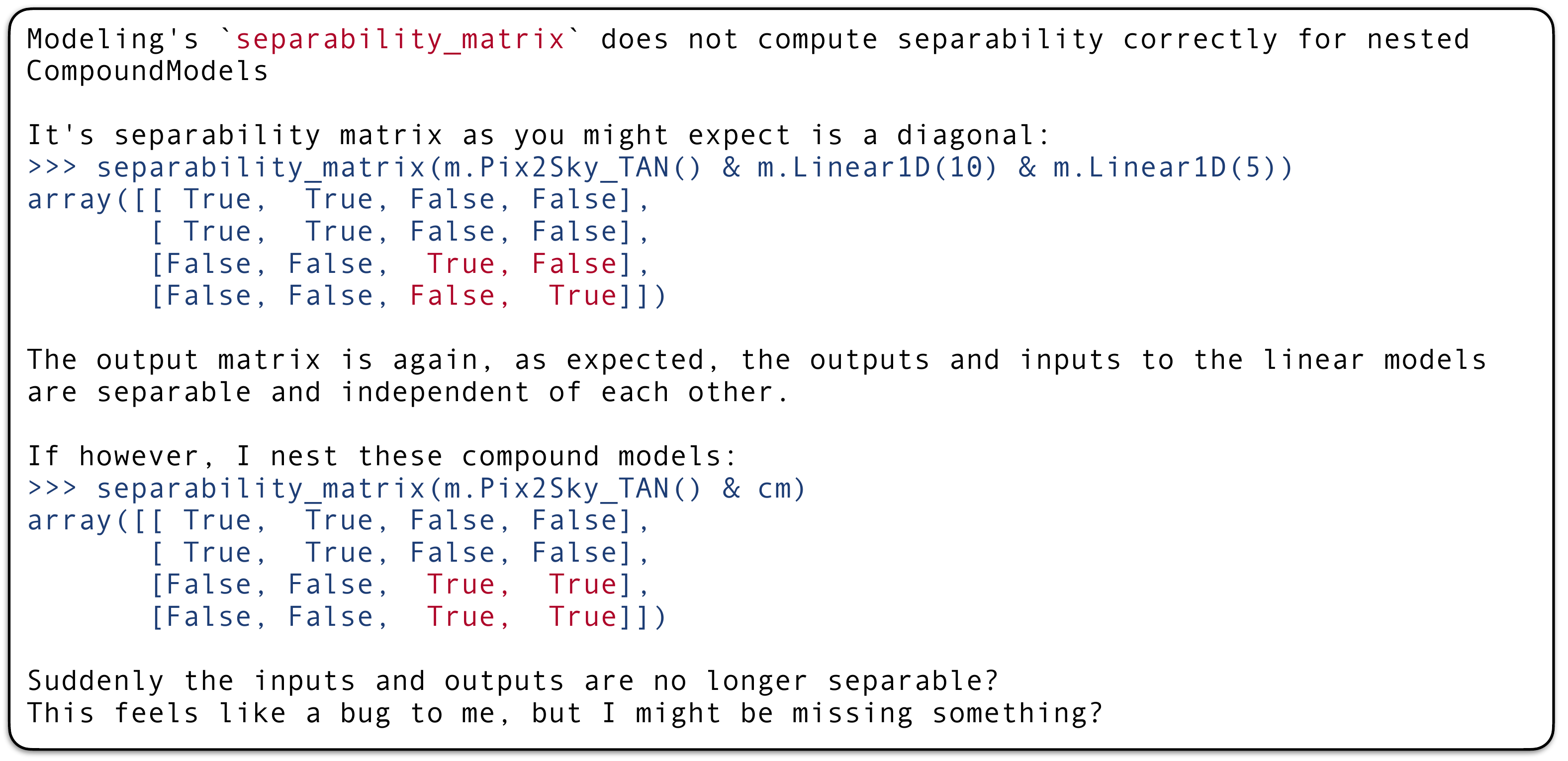}
    \caption{A real-world automated program repair task from the SWE-bench (\texttt{astropy-12907}~\cite{astropy12907})}
    \label{fig:astropy-12907-issue}
\end{figure}


This section presents a motivating example using a real-world automated program repair task from the \swebench{} dataset, \texttt{astropy-12907}~\cite{astropy12907} (with its issue description shown in Figure~\ref{fig:astropy-12907-issue}). 
Specifically, the task aims to fix the unexpected behavior of the \texttt{separability\_matrix} function in the Astropy~\cite{astropyRepo} library, which determines if a model's inputs and outputs are independent. While the function behaves as expected for simple composite models, it produces an incorrect result for nested composite models (colored in red), falsely indicating that their inputs and outputs are entangled. This task is challenging as it is a silent logical error with no clear failure signal, requiring an agent to diagnose it not by the final output, but by inspecting the intermediate states of the recursive calculation process to form hypotheses for the program's dynamic behavior. 


\begin{figure}[htb]
    \centering
    \includegraphics[width=0.92\columnwidth]{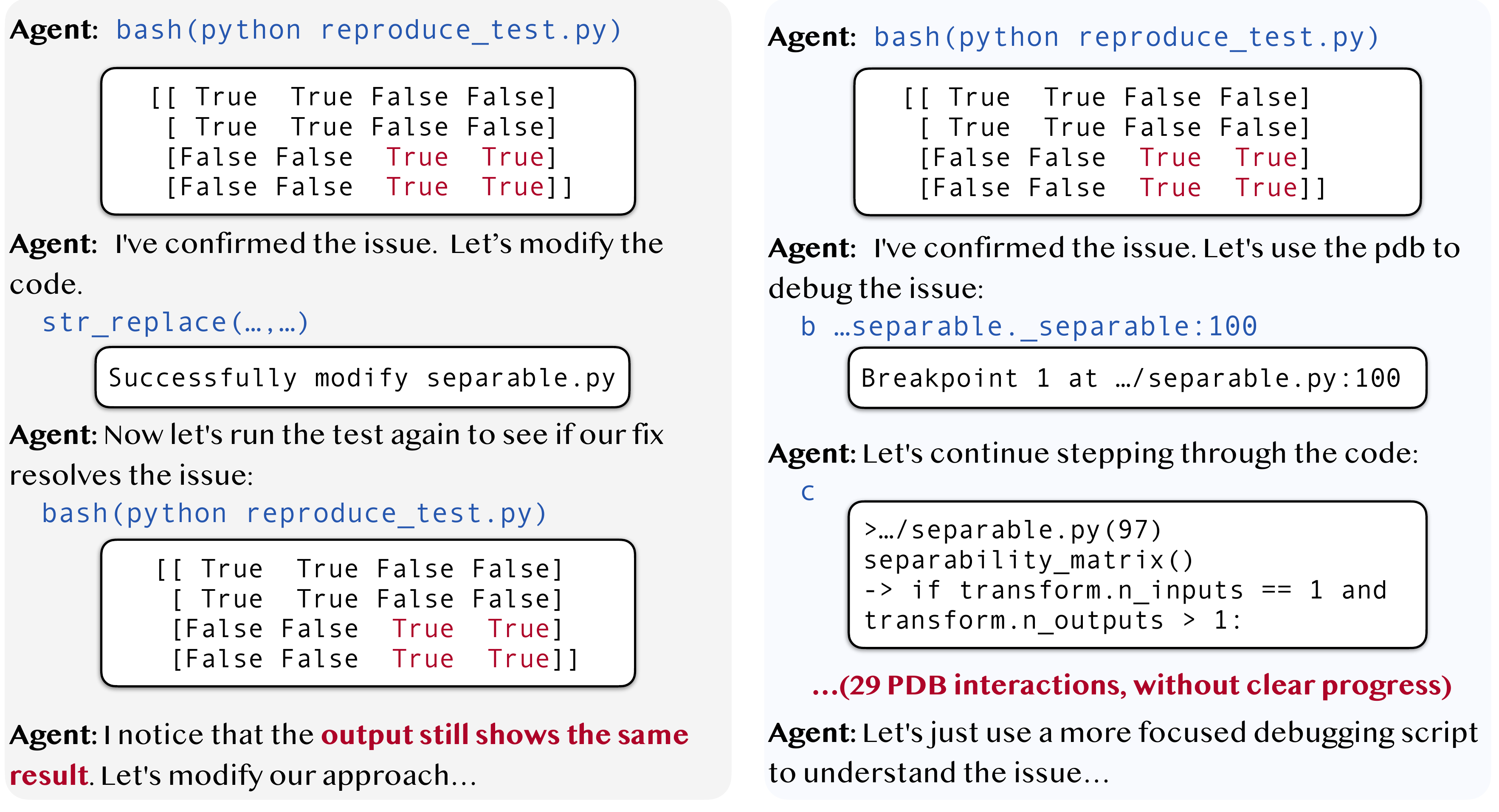}
    \caption{\baseagent{}'s and \pdbagent{}'s debugging processes on the \texttt{astropy-12907} task.}
    \label{fig:astropy-12907-base-pdb}
\end{figure}


To solve this task, we first employ an agent with the advanced LLM \claudesonnet{} that performs post-mortem debugging using only final execution outputs. As shown in Figure~\ref{fig:astropy-12907-base-pdb} (left), this agent halts in the cycle of reasoning and repeated test executions. Specifically, after confirming the issue and applying a blind patch, it re-runs the test only to observe that the ``output still shows the same result''. This indicates that having little information about the intricate execution path prevents the agent from understanding the root cause and leads to generating failed patches.

\begin{figure}[!h]
    \centering
    \includegraphics[width=0.95\columnwidth]{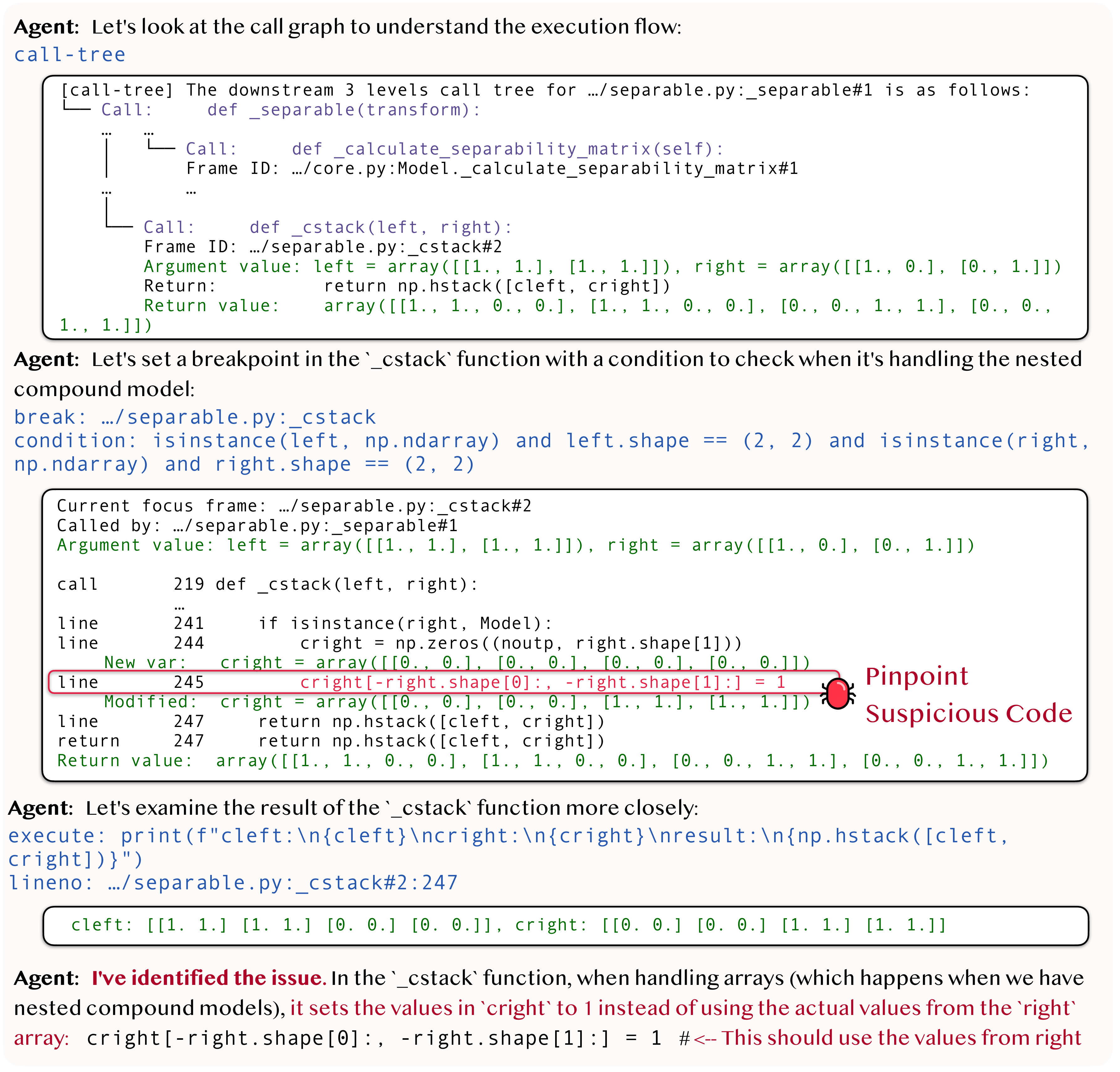}
    \caption{An Efficient Debugging Process on the \texttt{astropy-12907} Task.}
    \label{fig:astropy-12907-dbg}
\end{figure}

Next, we equip an agent with a standard PDB interface to issue low-level, atomic commands. However, as illustrated in Figure~\ref{fig:astropy-12907-base-pdb} (right), this approach introduces a severe efficiency bottleneck. The agent is forced to issue a long sequence of low-level commands (e.g., \texttt{b (break)}, \texttt{n (next)}) simply to concentrate on the suspicious code region. This process is proven highly inefficient, as each command triggers a costly LLM inference while yielding only a fragment of the program's state, eventually wasting the agent's effort after 29 unproductive rounds of interactions.

Ideally, an autonomous agent could resolve this task with the high efficiency and precision illustrated in Figure~\ref{fig:astropy-12907-dbg}. Rather than the tedious line-by-line stepping, such an agent would first capture a high-level summary of the top-level \texttt{\_separable} function's entire execution. By analyzing this summary and the corresponding call tree, the agent can deduce that while correct data flows into the nested \texttt{\_cstack} function, the matrix it returns is corrupted, thus pinpointing it as the source of the error. To confirm this hypothesis, the agent would then bypass irrelevant execution paths by setting a conditional breakpoint on \texttt{\_cstack}. This enables the inspection precisely at the suspicious code region, where a targeted query provides unambiguous evidence of the root cause: the \texttt{cright} array is incorrectly filled with ones. With this confirmation, the agent can formulate the correct patch, resolving the issue with the same logic as the official developer-written solution. \cmr{This motivating example illustrates the} value of a well-designed agent-debugger interface.

%% file: background.tex
\section{Background \& Related Work}

\subsection{Traditional Debuggers \& Function Frame}

\begin{figure}[htb]
    \centering
    \includegraphics[width=0.93\columnwidth]{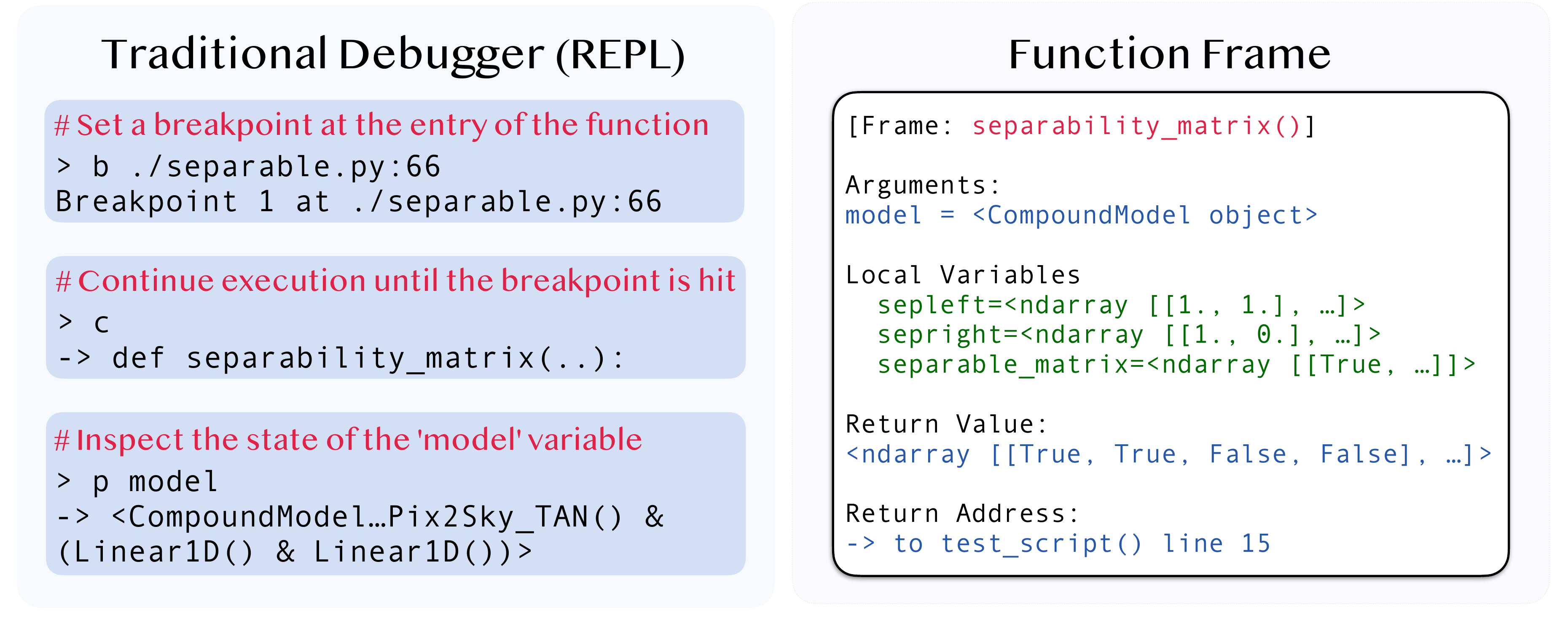}
    \caption{Traditional Debugger REPL Interaction and the Function Frame.}
    \label{fig:traditional-debugger}
\end{figure}

Traditional interactive debuggers, such as GDB~\cite{gdb} and PDB~\cite{pdb}, provide a powerful, human-in-the-loop interface for investigating program behaviors. As in Figure~\ref{fig:traditional-debugger} (left), their designs are based on a Read-Eval-Print Loop (REPL) where a developer issues a sequence of low-level, atomic commands like \texttt{step}, \texttt{next}, and \texttt{print} to trace a program's execution path with one statement at a time. 

The underlying data structure that enables this statement-level inspection of the program’s state is the program's call stack which tracks all active function calls. Each time a function is invoked, the corresponding function frame, i.e., activation record, is pushed onto the top of the stack. As depicted in Figure~\ref{fig:traditional-debugger} (right), this frame encapsulates the essential execution context for that specific function call, containing its arguments, local variables, and the return address to its caller. The associated Last-In, First-Out (LIFO) structure indicates that the stack of frames creates a nested record of the execution path, with the top frame always representing the ongoing executing function. Such a design reveals that the function frame is essentially a natural, higher-level unit of abstraction, potentially representing a function's execution context. This insight positions the function frame as the \cmr{natural} focal point for an agent-centric debugging interface.

\subsection{Agent-based Debugging}


Researchers have explored various approaches to enhance the debugging capabilities of LLM-based agents, with a prominent focus on iterative self-correction from execution feedback. Specifically, Self-Debugging~\cite{selfdebug} teaches an agent to perform rubber duck debugging by comparing natural language explanations of its code against the execution outcome. Similarly, ChatRepair~\cite{chatrepair} establishes a conversational repair loop where the agent learns from test failures to generate a more accurate patch. RepairAgent~\cite{bouzenia2024repairagent} further extends this by treating the LLM as an autonomous agent that freely decides which tool to invoke next, interleaving actions like gathering information and attempting a fix based on feedback from prior attempts. Nowadays, the idea of self-correction established by these works has become the primary debugging paradigm for many state-of-the-art autonomous agents~\cite{sweagent, openhands, bouzenia2024repairagent, specrover}. \cmr{Related work has also explored adjacent directions such as execution-trace-driven repair~\cite{bouzenia2023tracefixer}, conversational debugging assistants~\cite{bajpai2024let}, and notebook-based error-resolution agents~\cite{grotov2024debug}.} Another group of work utilizes fault localization (FL) techniques to narrow down the search space for a bug before a repair is attempted, e.g., agents like \acr{}~\cite{specrover} can optionally leverage Spectrum-Based Fault Localization (SBFL)~\cite{sbfl} to prioritize searching within suspicious code regions identified by test execution coverage. However, the effectiveness of these approaches is limited by their core dependencies, i.e., self-correction on post-mortem, coarse-grained execution output and fault localization on precise trigger tests, which prevent the agent from interactively exploring the program's dynamic state space.

To overcome the above-mentioned limitations of post-mortem, coarse-grained debugging analysis, one alternative solution is to equip agents with interactive debuggers to access fine-grained, dynamic program state. For instance, ChatDBG~\cite{chatdbg} enables a collaborative dialogue where an LLM agent autonomously controls standard debuggers like GDB and PDB to investigate a programmer's high-level queries. To facilitate the development of such agents, debug-gym~\cite{debuggym} provides a text-based interactive environment for agents to learn how to use debugging tools for code repair. Building on this, other works have applied interactive debugging to specific domains or with structured methodologies. EnIGMA~\cite{enigma} introduces Interactive Agent Tools to wrap command-line utilities like GDB for solving complex cybersecurity tasks. AutoSD~\cite{autosd} employs a scientific debugging approach, where the agent forms hypotheses and then uses a debugger to conduct experiments to validate them. Similarly, VulDebugger~\cite{vuldebugger} treats debugging as a state-comparison problem, using a debugger to continuously compare the program's actual runtime state against an expected correct state to repair vulnerabilities. However, simply integrating these human-centric debuggers with automated software engineering agents introduces an inherent mismatch. These debuggers operate on a Read-Eval-Print Loop (REPL), requiring a sequence of low-level, atomic commands (e.g., next, step, print var). For an autonomous agent, this interaction model is highly inefficient, where the costly LLM inference of each interaction round on a low-level command yields only a fragment of the program's state, eventually making the debugging process overwhelmingly expensive and compromising the agent's power on autonomously fixing complex bugs in an end-to-end manner. As a result, there is an urgent need for agent-centric interactive debugging interfaces which can facilitate the efficient retrieval of semantically rich, high-level summaries of program state, aligning the debugging process with the inferential capabilities of LLM agents.

%% file: approach.tex
\section{Agent-centric Debugging Interface} 

\begin{figure}[htb]
    \centering
    \includegraphics[width=0.97\columnwidth]{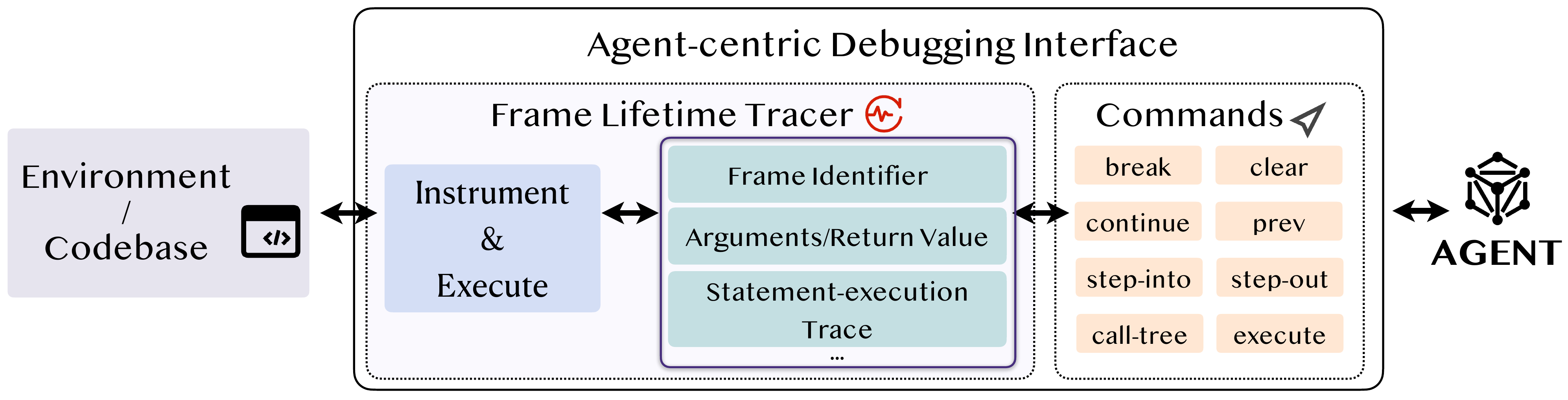}
    \caption{\adi{} Framework}
    \label{fig:adiframework}
\end{figure}

In this section, we introduce \adi{} (\adis{}) to realize cost-efficient
end-to-end autonomous function-level agent-debugger interaction, freeing agents from reliance on human-in-the-loop guidance. 
Figure~\ref{fig:adiframework} provides an overview of this framework, where a \textit{Frame Lifetime Tracer} instruments and executes a codebase to generate a series of \textit{\flt{}}s (FLTs). Each FLT is a self-contained data structure that captures the complete lifetime of a single function invocation, with its arguments, return value, and a detailed trace of all internal state modifications. An agent then inspects and navigates using a set of high-level \textit{Commands}. In particular, we model the interactive debugging session as a state transition system. The states in this system are defined over our core data abstraction, i.e., the \flt{} (FLT). The transitions are triggered by a set of agent-centric commands with well-defined semantics. These components are unified by the \adis{} Interaction Algorithm to realize the agent-driven dynamic analysis for automated program repair. \cmr{Importantly, \adis{} does not pre-compute a complete statement-level trace for an entire execution. Instead, it first records a lightweight function-level frame sequence and constructs a detailed \flt{} only for the frame currently selected by the inspection index.}


\subsection{Frame Lifetime Trace}

The core abstraction of \adis{} is the \flt{} (FLT). It is a self-contained data structure that captures the complete lifetime of a single function invocation, from its entry to its exit. This allows the agent to reason about a function's behavior holistically, rather than statement by statement.
Formally, an FLT is defined as a tuple:

\[ FLT = (\mathit{fid}, \mathit{fid}_\mathit{caller}, args, ret, \tau) \]

Where each component is defined as:
\begin{itemize}[leftmargin=*]
    \item \textbf{$\mathit{fid}$ (Frame Identifier):} A unique identifier for the target function invocation. It is constructed as a pair \texttt{(func, index)}, where \texttt{func} is the fully-qualified~\cite{Python3Glossary} function name and \texttt{index} is an invocation index. For example, \texttt{(my\_class:my\_func, 3)} refers to the invocation of \texttt{my\_func} during the execution for the third time.
    \item \textbf{$\mathit{fid}_\mathit{caller}$ (Caller Frame Identifier):} The frame identifier ($\mathit{fid}$) of this frame's immediate caller. This is crucial for navigating the call stack. For the entry-point frame, this can be \texttt{null}.
    \item \textbf{$args$ (Arguments):} A key-value map of the function's formal parameter names to their corresponding values at the snapshot of invocation.
    \item \textbf{$ret$ (Return Value):} The value returned by the function upon its exit. If the function terminates upon an unhandled exception, this component may contain exception information.
    \item \textbf{$\tau$ (Statement-execution Trace):} An ordered trace of all statements executed within the function's lifetime. The trace is a sequence of \textit{execution steps}, where each step is formally defined as the tuple:
    \[
        step = (lineno, stmt, \Delta S, \mathit{fid}_\mathit{callee})
    \]
    \begin{itemize}[leftmargin=*]
        \item \textbf{$lineno$} (Line Number): The line number of the executed statement in the source code.
        \item \textbf{$stmt$} (Statement): The source code of the statement itself.
        \item \textbf{$\Delta S$} (State Modifications): A set of state modifications caused by this statement. Each modification is defined as a tuple \texttt{(variable, old\_value, new\_value)}, capturing the definition of new variables or changes to existing ones. If a statement causes no state change, $\Delta S$ is empty.
        \item \textbf{$\mathit{fid}_\mathit{callee}$} (Callee Frame Identifier): If the statement invokes a function, this field contains the identifier ($\mathit{fid}$) of the resulting frame (the callee). Otherwise, it is \texttt{null}.
    \end{itemize}
\end{itemize}

\begin{figure}[htb]
    \centering
    \includegraphics[width=0.92\columnwidth]{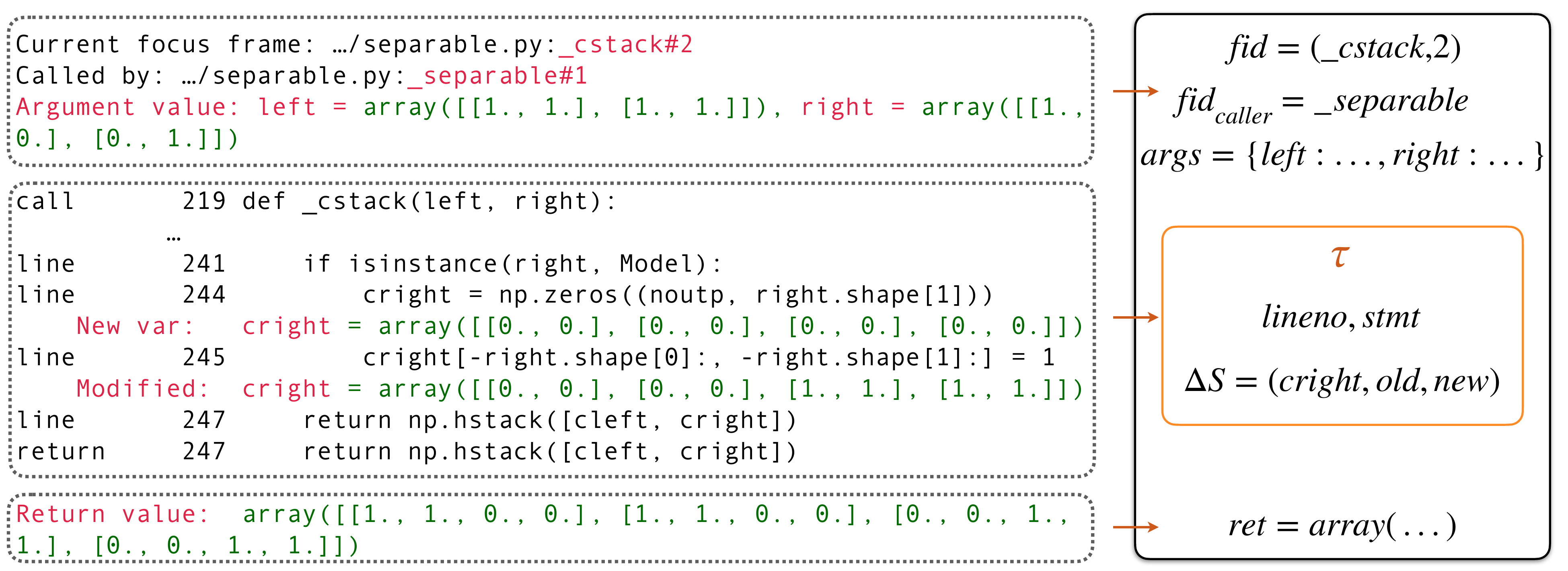}
    \caption{The \flt{} (FLT) of the \texttt{\_cstack\#2} invocation from the \texttt{astropy-12907} task}
    \label{fig:flt-12907}
\end{figure}

Figure~\ref{fig:flt-12907} presents an FLT instance for the \texttt{\_cstack\#2} invocation from our motivating example, illustrating the mapping between its textual representation (left) and its formal components (right). 
Note that the function's representative invocation context, including its caller and arguments, is mapped to the $\mathit{fid}$, $\mathit{fid}_\mathit{caller}$, and $args$ components. 
The statement-execution trace ($\tau$) records the execution sequence and presents the resulting state modifications for each statement ($\Delta S$).
 For instance, the sequence shows the \texttt{cright} variable being created (\texttt{New var}) at line 244 and subsequently modified (\texttt{Modified}) at line 245. Finally, the function's exit value is encapsulated in the $ret$ component. 



We then model the interactive debugging session as a state transition system based on the FLT concept. 
The complete state of a debugging session is captured by a tuple $\Sigma$:
\[
\Sigma = (B, T, i)
\]
with the parameters illustrated as follows. 
\begin{itemize}[leftmargin=*]
    \item \textbf{$B$ (Breakpoint Set):} A set of pairs $(\mathit{func}, \mathit{condition})$, where \texttt{func} is a function identifier and \texttt{condition} is an optional boolean expression evaluated at the function entry (by default \texttt{True}).
    \item \textbf{$T$ (Full Execution Trace):} An ordered sequence $\langle FLT_0, FLT_1, \ldots, FLT_n \rangle$ representing all function calls from a single program execution, ordered by invocation time.
    \item \textbf{$i$ (Inspection Index):} An index $i \in \{0, \ldots, n\}$, that identifies the ongoing frame $FLT_i$ under the agent's inspection. 
\end{itemize}

Then, the commands in the following section are defined as functions that map a state $\Sigma$ to a new state $\Sigma'$, denoted $\Sigma \to \Sigma'$.

\subsection{Agent-Centric Debugging Commands}
\label{sec:adicmd}


The \adis{} command set equips the agent with a structured interface to explore the program's execution trace $T$. 
First, the agent uses \texttt{break} and \texttt{clear} to dynamically manage a set of function-level conditional breakpoints, defining which execution frames are points of interest. Once these points are established, the agent can navigate the program's dynamic state. The \texttt{continue} and \texttt{prev} commands enable chronological traversal along the execution timeline, moving between breakpoint hits, while \texttt{step-out} and \texttt{step-into} serve as their respective reverse operations to navigate the call stack between caller and callee frames. This bidirectional navigation allows the agent to freely access the execution history. Finally, for deeper inspection without altering its position, the agent can use query commands: \texttt{call-tree} provides a local overview of the call hierarchy, and \texttt{execute} facilitates ``what-if'' analysis by running a statement under a specific context. Each command transforms the debugging state $\Sigma$ by either updating the breakpoint set $B$ or the inspection index $i$. 

\begin{itemize}[leftmargin=*]
\item \textbf{\texttt{break(func, [condition])}:}
Adds a breakpoint to the set $B$.
\cmr{The optional \texttt{condition} is a boolean expression evaluated at the function entry or exit, with the default value \texttt{True}.}
The state transition is
\[
\Sigma \to \Sigma' 
\quad \text{where } 
\Sigma' = (B \cup \{ (\mathit{func}, \mathit{condition})\}, T, i).
\]

This command allows the agent to register specific function invocations as points of interest for subsequent navigation. Specifically, this operation updates the state solely by adding the new \texttt{(func, condition)} pair to the breakpoint set $B$.

\item \textbf{\texttt{clear(func, [condition])}:} 
The inverse command of \texttt{break}, which removes a specified breakpoint from the set $B$.

\item \textbf{\texttt{continue()}:} Advances the inspection index forward to the next frame that matches a registered breakpoint. In particular, we set $i$ as the current inspection index. First, the set $J$ collects the indices $j$ of all subsequent frames ($j > i$) that match a breakpoint. Then the new index $i'$ is set to the first of these matches ($\min(J)$), or remains unchanged if no future match exists. 
\[
\Sigma \to \Sigma' 
\quad \text{where } 
\Sigma' = (B, T, i'),
\]
with $i'$ determined as:
\[
\begin{aligned}
J &= \{j \mid j > i \land (T[j].\mathit{fid}.\mathit{func}, \, \mathit{condition}) \in B \land \operatorname{evaluate}(T[j].\mathit{args}, \mathit{condition})\}, \\
i' &=
\begin{cases}
\min(J) & \text{if } J \neq \varnothing, \\
i       & \text{otherwise}.
\end{cases}
\end{aligned}
\]

After executing this command, the agent receives the FLT $T[i']$.

\item \textbf{\texttt{prev()}:} 
The inverse command of \texttt{continue}, which moves the inspection index backward to the most recent preceding breakpoint match and returns the corresponding frame $T[i']$.

\item \textbf{\texttt{step-into(fid)}:} 
This command provides targeted navigation, moving the inspection index directly to a frame specified by its unique identifier (\texttt{fid}). It allows the agent to ``step into'' a specific function call it has identified, such as a callee discovered from the current context, without needing to traverse the execution timeline sequentially. Specifically, this operation locates the frame with the matching \texttt{fid} within the entire trace $T$ and updates the current inspection index $i$ to the found frame's index $j$; the index remains unchanged if the \texttt{fid} is not found.
The state transition is
\[
\Sigma \;\to\;\; \Sigma' 
\quad \text{where } 
\Sigma' = (B, T, i'),
\]
with $i'$ determined as:
\[
i' =
\begin{cases}
  j & \text{if } \exists j,\; T[j].\mathit{fid} = \mathit{fid}, \\
  i & \text{otherwise}.
\end{cases}
\]
After executing this command, the agent receives the FLT $T[i']$.

\item \textbf{\texttt{step-out()}:} 
The inverse command of \texttt{step-into}, which moves the inspection index up the call stack from the current frame to its direct caller, returning the corresponding frame $T[i']$.


\item \textbf{\texttt{call-tree()}:} 
This query command returns a tree-like summary of the downstream call hierarchy from the current frame, allowing the agent to preview subsequent execution paths. Each node in the tree represents a function call, annotated with its signature, \texttt{fid}, arguments, and return value. This command is read-only and does not change the debugging state ($\Sigma' = \Sigma$).


\item \textbf{\texttt{execute(fid, stmt, lineno, k)}:} 
This command provides a powerful ``what-if'' analysis capability by dynamically executing a statement within a specific execution context. It is designed for fine-grained state inspection, such as checking a variable's value on a particular loop iteration without setting a complex breakpoint.
\cmr{Upon invocation, the command re-executes the program to reach the target frame identified by \texttt{fid} and the $k$-th visit to line \texttt{lineno}. It then temporarily injects \texttt{stmt} in that execution context and returns the resulting output. This operation does not modify the source code or the canonical trace $T$.}
\end{itemize}

\subsection{ADI Interaction Algorithm}

\label{sec:adi-algorithm}

\input{figures/adi-algo}

Algorithm~\ref{alg:adi} presents a detailed procedure that integrates \flt{}s with agent-driven interaction. It consists of two cooperating components: the \textsc{FrameLifetimeTracer} and the \textsc{ADI\_Interaction\_Loop}. The \textsc{ADI\_Interaction\_Loop} serves as the main engine that processes the interactive navigation commands defined in Section~\ref{sec:adicmd}, while the \textsc{FrameLifetimeTracer} is a specialized procedure invoked on-demand to construct a \flt{}. 

Specifically, \textsc{FrameLifetimeTracer} constructs a \flt{} for a single function invocation by applying statement-level instrumentation to the selected frame. When the function is invoked, it records the frame identifier (\(\mathit{fid}\)), the caller frame identifier  (\(\mathit{fid}_\mathit{caller}\)), and the arguments (\(args\)) (lines 2-4). During the lifetime of the frame, the procedure repeatedly processes the statements executed in the function body (line 6). The \textsc{NextStmt} returns the next statement together with its line number \((stmt,\ \ell)\). Before and after executing the statement, \textsc{RecordEnv} captures the ongoing variable environment (lines 7 and 9). The difference between the two environments yields the state modification \(\Delta S\) (line 10), which represents the updates caused by the statement. The routine \textsc{ExecStmt} executes the current statement, returning the identifier of a callee frame (\(\mathit{fid}_\mathit{callee}\)) if the statement invokes a function, and updates the exception variable \(\epsilon\) if an unhandled exception occurs. 
Each execution step is thus represented as a tuple \((\ell, stmt, \Delta S, \mathit{fid}_\mathit{callee})\) and appended to the statement-execution trace \(\tau\). 
When the frame terminates, if \(\epsilon \neq \texttt{null}\), the return field is set to the exception information; otherwise, it is set to the normal return value (lines 14-17). As a result, an FLT 
\((\mathit{fid}, \mathit{fid}_\mathit{caller}, args, ret, \tau)\) summarizing the invocation is completed (line 18).

The procedure \textsc{ADI\_Interaction\_Loop} in Algorithm~\ref{alg:adi} manages the interactive debugging session. It begins by executing the program under lightweight function-level instrumentation to initialize the sequence of $\mathit{fid}$s (line 20). The debugging state is initialized as \(\Sigma = (B=\emptyset,\ T=T,\ i=0)\), where $B$ is the breakpoint set, $T$ is the frame sequence, and $i$ is the inspection index (line 21). The loop then repeatedly accepts commands from the agent (line 22). Each command is processed according to the state-transition rule of Section~\ref{sec:adicmd}, yielding an updated state \(\Sigma'\) (lines 23-24). From the new inspection index \(i\), \textsc{ADI\_Interaction\_Loop} retrieves the identifier \(\mathit{fid}_\mathit{insp} = T[i].\mathit{fid}\) and invokes \textsc{FrameLifetimeTracer}\((\mathit{fid}_\mathit{insp})\) (lines
 25-26) to construct the corresponding FLT. It then derives the response such as the new inspection frame based on \(\Sigma\) and the constructed FLT, and returns it to the agent (lines 27-28). With such a loop, the agent issues a navigation command, and in response, the \adis{} dynamically constructs a rich, detailed view (the FLT) of the new inspection frame. This on-demand analysis ensures that the agent receives deep contextual information precisely when and where it is needed for the next decision. 

%% file: figures/adi-algo.tex
\begin{algorithm}[t]
\caption{ADI Interaction Algorithm}
\label{alg:adi}
\begin{algorithmic}[1]
\Require Program $P$, Input data $I$

\Statex
\Function{FrameLifetimeTracer}{$\mathit{fid},\ P,\ I$}
  \State $\mathit{flt} \gets \Call{NewFLT}{\mathit{fid}}$
  \State $\mathit{flt}.\mathit{fid}_{caller} \gets \Call{CallerFID}{}$
  \State $\mathit{flt}.args \gets \Call{ArgsAtEntry}{}$
  \State $\epsilon \gets \texttt{null}$ \Comment{Initialize exception info}
  \While{$(stmt,\ \ell) \gets \Call{NextStmt}{\mathit{fid}}$} \Comment{Statement-level instrumentation}
    \State $Env_\mathit{before} \gets \Call{RecordEnv}{}$
    \State $(\mathit{fid}_{callee},\ \epsilon) \gets \Call{ExecStmt}{stmt,\ P,\ I}$
    \State $Env_\mathit{after} \gets \Call{RecordEnv}{}$
    \State $\Delta S \gets \Call{DeltaState}{Env_\mathit{before},\ Env_\mathit{after}}$ \Comment{Compute state modifications}
    \State $\mathit{flt}.\tau.\Call{Append}{(\ell,\ stmt,\ \Delta S,\ \mathit{fid}_{callee})}$
    \If{$\epsilon \neq \texttt{null}$} \State \textbf{break} \EndIf
  \EndWhile
  \If{$\epsilon \neq \texttt{null}$}
    \State $\mathit{flt}.ret \gets \epsilon$ \Comment{Return exception info}
  \Else
    \State $\mathit{flt}.ret \gets \Call{ReturnValue}{}$
  \EndIf
  \State \Return $\mathit{flt}$
\EndFunction

\Statex
\Function{ADI\_Interaction\_Loop}{$P,\ I$}
  \State $T \gets \Call{BuildFrameSequence}{P,\ I}$ \Comment{For efficiency, only sequence of $\mathit{fid}$s is collected}
  \State $\Sigma \gets (B=\varnothing,\ T=T,\ i=0)$
  \While{$cmd \gets \Call{ReceiveCmd}{}$}
    \State $\Sigma' \gets \Call{ApplyTransition}{\Sigma,\ cmd}$ \Comment{Update debugging state}
    \State $\Sigma \gets \Sigma'$
    \State $\mathit{fid}_\mathit{insp} \gets T[i].\mathit{fid}$
    \State $\mathit{flt}_\mathit{insp} \gets \Call{FrameLifetimeTracer}{\mathit{fid}_\mathit{insp},\ P,\ I}$ \Comment{Construct FLT of the inspection frame}
    \State $out \gets \Call{DeriveResponse}{cmd,\ \Sigma,\ \mathit{flt}_\mathit{insp}}$
    \State \Call{SendToAgent}{$out$}
  \EndWhile
\EndFunction

\end{algorithmic}
\end{algorithm}

%% file: evaluation.tex
\section{Evaluation}

In this section, we investigate the effectiveness and characteristics of \adi{} (\adis{}). In particular, we build an agent equipped with \adis{}, namely \dbgagent{}. We attempt to answer the following research questions:

\begin{itemize}[leftmargin=*]
    \item \parabf{RQ1:} \textit{How effective is \adis{} in advancing automated program repair?} To answer this RQ, we evaluate \dbgagent{} and compare its performance against other baselines.
    \item \parabf{RQ2:} \textit{How do the state-of-the-art agents equipped with \adis{} perform?} To answer this RQ, we integrate \adis{} into \sweagentmini{}~\cite{sweagent} and \acr{}~\cite{specrover} and compare the performance of these enhanced agents against their original versions to evaluate \adis{}'s generality as a plug-and-play enhancement.
    \item \parabf{RQ3:} \textit{What are the behavioral characteristics of \adis{}?} To answer this RQ, we analyze the behavioral characteristics of \dbgagent{} to understand how \adis{} shapes its problem-solving strategies.

\end{itemize}

\subsection{Benchmark and Evaluation Metrics}

We adopt the popular \swebench{} dataset~\cite{swebench} to assess how the \adi{} enhances an agent's performance on real-world automated program repair tasks. In particular, we focus on the widely-used \swebench{} Lite~\cite{swebench} and Verified~\cite{OpenAI2024SWEBench} versions, containing 300 and 500 tasks, respectively. These benchmarks employ developer-written unit tests to verify the correctness of agent-generated patches, ensuring a rigorous assessment of the agent’s performance.

Following prior work~\cite{agentless, specrover, sweagent}, our major evaluation metric is the \textbf{Resolved Rate (\%)}, reported as \textbf{Pass@1} efficacy. A task is considered resolved if a single generated patch successfully applies to the codebase and passes all developer-written acceptance tests. Crucially, these acceptance tests are held out and not used by the agent during the patch generation process to ensure a fair evaluation. Specifically, to ensure consistency and reproducibility, all evaluations are conducted using the official SWE-bench Docker environment provided by the SWE-bench team~\cite{swebench_website}. Moreover, we report the average API inference cost (\textbf{\$ Avg. Cost}) and \textbf{Correct Location Rate (\%)}, i.e., the percentage of generated patches covering the ground-truth edit locations at the file, hunk, and line granularities~\cite{agentless, swegpt}.

\subsection{Baselines and LLMs}

\paragraph{Baselines} We first adopt six state-of-the-art agents and approaches as our baselines: \ct{}~\cite{anthropic2025claude3_7}, \sweagent{}~\cite{sweagent}, \sweagentmini{}~\cite{swebench_website}, \acr{}~\cite{specrover}, \ssi{}~\cite{swegpt}, and \agentless{}~\cite{agentless}. In particular, we primarily utilize the performance results reported in their original publications~\cite{agentless, sweagent, specrover, swegpt} following~\cite{swegpt} and the official \swebench{} leaderboard~\cite{swebench}. For our generality evaluation in RQ2, we integrate \adis{} into two agents selected for their architectural diversity and suitability as integration targets. First, we select \sweagentmini{} as a representative of the ReAct~\cite{react}-based agents. Its lightweight and open design makes it an ideal testbed for integrating and evaluating components like \adis{}. In contrast, \acr{} employs a fundamentally different retrieve-and-generate strategy. This selection allows us to verify whether the benefits of \adis{} are not confined to a single agent design but can serve as a general-purpose enhancement across diverse agent architectures.


We also design two variants of \dbgagent{} for further ablation studies on \adi{}: \baseagent{} and \pdbagent{}. 
Specifically, \baseagent{} represents a standard \textit{post-mortem debugging} approach and equips only with basic file editing and execution tools.  \pdbagent{} extends \baseagent{} with a conventional PDB interface, representing a naive integration of \textit{human-in-the-loop interactive debugging}.

\paragraph{LLMs} We adopt four advanced LLMs: \claudesonnet{} (claude-3-7-sonnet-20250219)~\cite{anthropic2025claude3_7}, \claudethreefive{} (claude-3-5-20241022)~\cite{anthropic2024claude3_5}, \gpto{} (gpt-4o-2024-11-20)~\cite{openai2024gpt4o}, and \qwen{} (qwen3-32b)~\cite{Qwen_Pricing}. We obtain the open-source model \qwen{} from Hugging Face~\cite{hugging-face-fc} and access \claudesonnet{}, \claudethreefive{}, \gpto{} through the APIs provided by Anthropic~\cite{anthropic_claude_docs_2025} and OpenAI~\cite{openai-api-fc}. Inference for the open-source model \qwen{} is conducted on servers with 128-core 2.6GHz AMD EPYC™ ROME 7H12 CPU, 512 GiB RAM, and eight NVIDIA A100 80GB GPUs, running Ubuntu 20.04.6 LTS, utilizing vllm~\cite{kwon2023efficient} inference framework.

\subsection{Implementation and Setups}
\subsubsection{\adi{}}

A key challenge in implementing \adi{} is the prohibitive performance overhead of full-program, statement-level tracing. To mitigate this, we leverage the \texttt{sys.settrace} facility in Python for selective instrumentation. Instead of tracing the entire program, we activate fine-grained, statement-level tracing on-demand, only for the single frame currently designated by the agent's inspection index, thus providing deep visibility precisely where needed, without incurring the cost of global instrumentation. Specifically, we measure the runtime overhead of our \textsc{FrameLifetimeTracer} on SWE-bench Verified, observing a modest increase in average execution time from 0.68s to 0.87s when executing the official failing tests with tracing on the buggy function frames. \cmr{Moreover, following the \swebench{} protocol~\cite{swebench_issue16_test_patch}, developer-written failing tests are held out, and the agent instead generates a reproduction script from the issue description for ADI-based debugging.}

We also realized several targeted optimizations. For frames containing large loops, our tracer intelligently captures only the first and last iterations, reporting the intermediate ones as a single ``skipped'' block to the agent. The \texttt{call-tree} command is also bounded, exploring only three levels of the callee hierarchy to prevent the agent from being overwhelmed with excessive data. Finally, to support the \texttt{execute} command, we implemented code injection via Python's \texttt{exec} function. Ensuring that the injected code's side effects are correctly reflected in the running program requires synchronizing the frame's local variables using the CPython API call \texttt{PyFrame\_LocalsToFast}~\cite{cpythondoc}, which we access through the \texttt{ctypes} library~\cite{ctypelib}. This specific mechanism is crucial for reliable state manipulation during a debugging session.

\subsubsection{\dbgagent{} and its variants}

\begin{figure}[htb]
    \centering
    \includegraphics[width=0.85\columnwidth]{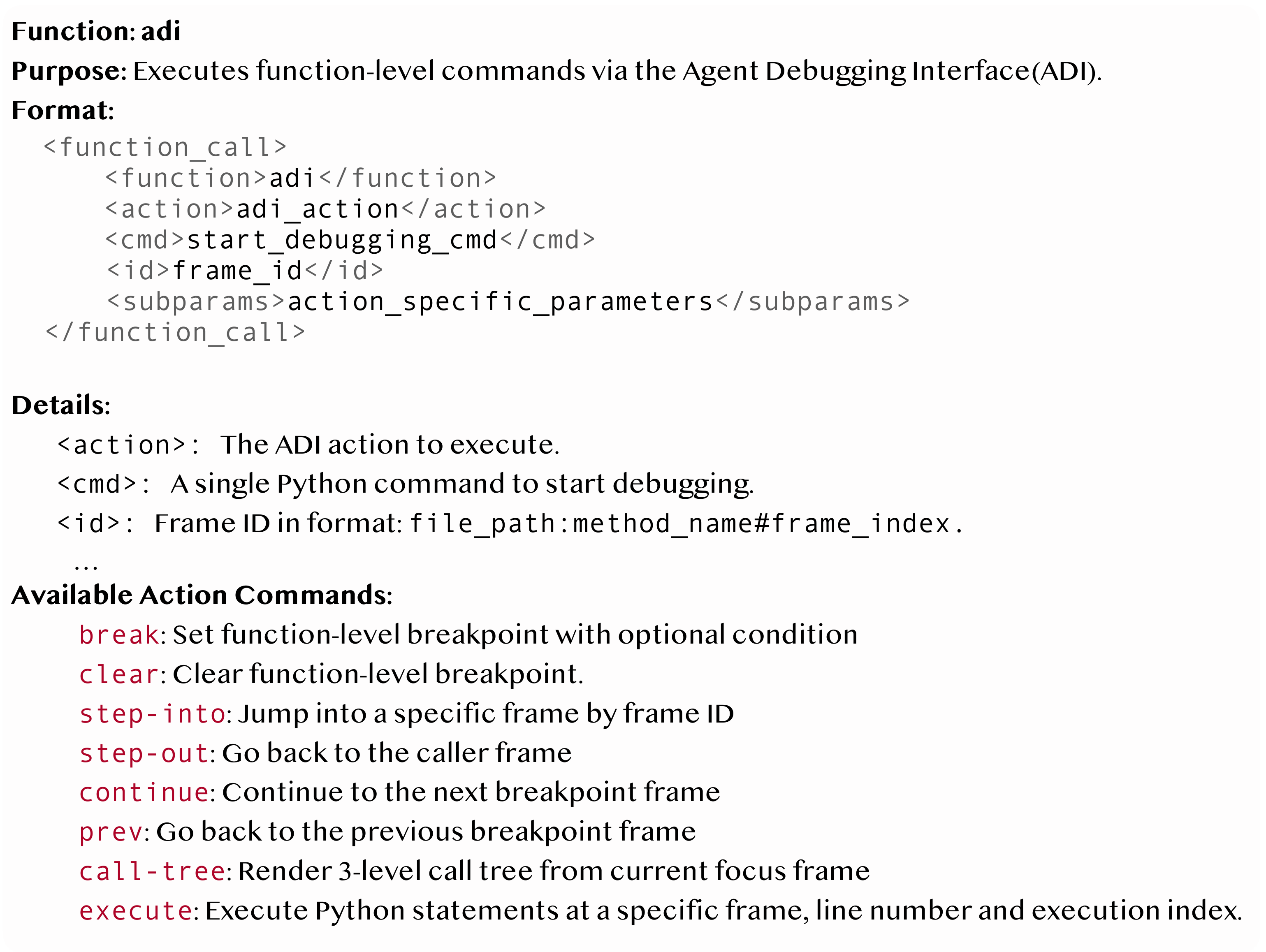}
    \caption{The prompt of \adis{} used by \dbgagent{}.}
    \label{fig:adi-prompt}
\end{figure}

\dbgagent{}, \baseagent{}, and \pdbagent{} are built upon the ReAct framework and the same workflow to ensure a fair comparison of their reasoning capabilities. In this framework, the LLM iteratively generates thoughts to reason about problems and actions to invoke tools, receiving observations from the system to form the next steps. \dbgagent{} automates tasks through a four-stage process: code orientation, issue reproduction, root-cause analysis, and patch implementation. Crucially, it leverages the function calls provided by our \adi{} to perform on-demand dynamic analysis, particularly during the analysis stage. In contrast, 
\baseagent{} is restricted to a standard toolkit for file operations and bash commands, relying solely on analyzing the final output of program executions. For \pdbagent{}, 
following prior work~\cite{debuggym, enigma}, we realize a non-blocking PDB session that the agent interacts with and manages through function calls. To illustrate how \adi{} is exposed to the agent, Figure~\ref{fig:adi-prompt} presents the detailed tool definition used by \dbgagent{}.


All three agents are evaluated under the same conditions. Following the \sweagent{} setup~\cite{sweagent}, we set a \$4 budget per task for cost control. Moreover, we follow the setup of baseline agents~\cite{sweagent, specrover}, with the model's temperature set to 0 to ensure that its output is more deterministic.

\subsubsection{Adapting SOTA Agents for \adis{}}

We integrated \adi{} into two SOTA agents, \sweagentmini{} and \acr{}, with minimal modifications to their native workflows. For \sweagentmini{}, we adapt its primary prompt with \adis{} usage instructions and introduce an optional dynamic analysis stage to its workflow, empowering it to invoke interactive debugging. For \acr{}, we enhanced its context retrieval agent by prompting it to use \adis{} for collecting richer, dynamic runtime information about program behaviors before attempting a patch. In both cases, the \adis{} toolset is deployed within the agents' sandboxed environments and exposed through a lightweight API, making it seamlessly available within their existing action spaces. Due to page limits, more implementation details are shown on our GitHub page~\cite{githubrepo}. 

\subsection{Result Analysis}

\subsubsection{RQ1: effectiveness of \adi{}}

\input{tables/baseline}

As shown in Table~\ref{tab:rq1-sota}, equipping a basic agent with \adis{} leads to highly competitive performance. When using \claudesonnet{} on the Verified benchmark, the \adis{}-enabled agent achieves a 63.8\% resolved rate. This result even slightly exceeds the performance of the highly-optimized and high-investment \tools{} agent (63.2\%), which serves as the foundation for Anthropic's commercial Claude Code product~\cite{anthropic2025claude3_7}. This strong performance extends to the \swebench{} Lite benchmark, where our agent also achieves a top-performing 50.7\% resolved rate. Notably, this leading performance is achieved at an average cost of only \$1.28 per task. Moreover, we observe that the \adis{}-enabled \dbgagent{} uniquely solves two tasks on the entire \swebench{} leaderboard~\cite{swebench_website} which none of the rest baselines could. All such results indicate that \adi{} is essentially powerful to advance the effectiveness of an agent for automated program repair tasks. 

\input{tables/rq1}

Table~\ref{tab:rq1-ablation-result} presents the results of our ablation study where we compare \dbgagent{} with \baseagent{} and \pdbagent{}. \dbgagent{} consistently outperforms both baselines across all LLMs and benchmarks. For instance, on \swebench{} Verified, \dbgagent{} achieves relative performance gains of 7.5\% to 16.0\% over \baseagent{} and 10.6\% to 16.0\% over \pdbagent{}. Similar gains are observed on \swebench{} Lite.

The performance gains from \adis{} are also highly cost-efficient. While \dbgagent{} incurs a modest cost increase over \baseagent{} (e.g., \$0.26 with \claudesonnet{}), the cost of \pdbagent{} is substantially higher (e.g., \$0.60). This inefficiency is rooted in the nature of traditional debugging. For instance, when using \claudesonnet{}, \pdbagent{} invokes the \texttt{next} command an average of 10.1 times in 191 tasks where it attempts debugging. Our manual analysis of all 191 of these attempts reveals that this inefficient process leads the agent to abandon its attempt in 53\% of them.

\mybox{Finding 1: \adi{} advances the capabilities of autonomous agents for automated program repair tasks. It enables a basic agent to achieve \cmr{superior} performance on \swebench{}, even slightly outperforming highly-optimized and high-investment tools while also solving tasks they cannot at a low cost.}


The strong performance and efficiency gains of \adis{} on a baseline agent motivate an inquiry into its generality. In RQ2, we then investigate whether \adis{} can serve as a plug-and-play enhancement for the existing state-of-the-art agents.

\subsubsection{RQ2: Generalizability to SOTA Agents}
\input{tables/rq3-adi_agent}

As shown in Table~\ref{tab:sota_comparison}, integrating \adi{} as a plug-and-play component consistently enhances the performance of both \sweagentmini{} and \acr{}. Across different LLMs, this integration delivers performance gains ranging from 6.2\% to 18.5\% at a modest cost increase. Specifically, for \sweagentmini{}, whose ReAct-based workflow relies on binary pass/fail signals, \adis{} provides a crucial enhancement. It equips the agent with the capability to inspect the program's internal state upon a failure, enabling targeted, diagnostic reasoning that goes beyond a simple test outcome. This enhanced reasoning capability boosts the number of resolved tasks by 10.6\% with \claudesonnet{} and 18.5\% with \gpto{}. Similarly, \adis{} augments \acr{}'s static analysis-based strategy by providing dynamic runtime information. This richer, hybrid dynamic context empowers the agent to generate more precise patches for bugs only observable at runtime, resulting in resolved rate improvements of 7.3\% with \claudethreefive{} and 6.2\% with \gpto{}. All such results demonstrate that \adis{} could enhance the performance of general agent architectures for automated program repair tasks.

\mybox{Finding 2: \adi{} acts as a general-purpose enhancement, consistently improving the performance of agents with distinct architectures at modest costs.}

\subsubsection{RQ3: characteristics of \adi{}}

\begin{figure}[htb]
    \centering
    \includegraphics[width=0.63\columnwidth]{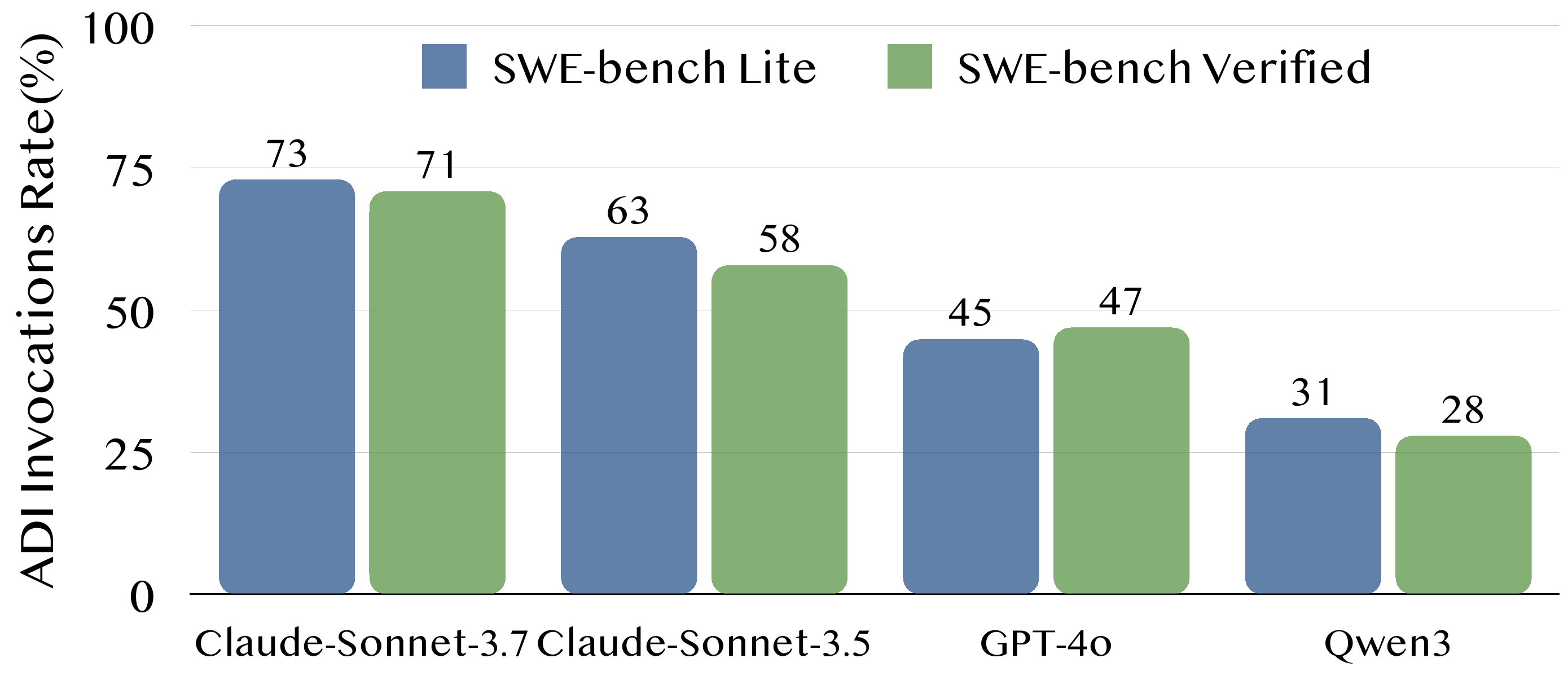}
    \caption{Invocation rate of \adis{} on the \swebench{} Lite and Verified benchmarks.}
    \label{fig:adi-call-times}
\end{figure}


We first investigate the characteristics of \adis{} by analyzing its invocation rate on the \swebench{} Lite and Verified benchmarks. Figure~\ref{fig:adi-call-times} reveals a strong correlation between the agent's utilization of the \adis{} and the capability of its underlying LLM. For instance, with \claudesonnet{}, the agent invokes \adis{} commands in 73\% of tasks on \swebench{} Lite and 71\% on \swebench{} Verified. In contrast, this rate drops significantly to just 31\% and 28\% for \qwen{}. Meanwhile, Table~\ref{tab:rq1-ablation-result} 
shows that on \swebench{} Verified, the resolved rate of \baseagent{} descends from 55.0\% for \claudesonnet{} to 29.2\% for \qwen{}. Such a trend aligns with the above-mentioned \adis{} invocation rates, indicating that stronger LLMs are more adept at leveraging \adis{}, allowing them to amplify their inherent analytical capabilities. 
\cmr{For the 62 tasks uniquely resolved by \dbgagent{} over \baseagent{} on \swebench{} Verified with \claudesonnet{}, ADI commands are invoked 6.39 times per task on average.}

Moreover, we find that \texttt{break} is the most frequently-used command, invoked an average of 2.5 times, followed by \texttt{continue} with an average of 2.4 invocations, across the four LLMs used by \dbgagent{}. Interestingly, when analyzing the \dbgagent{} with \claudesonnet{}, we observe that the \texttt{execute} command is invoked in 44 tasks with an average of 3.1 invocations each, while the \texttt{call-tree} command is used in 16 tasks, averaging 1.0 invocations. 

\begin{figure}[htb]
    \centering
    \includegraphics[width=0.88\columnwidth]{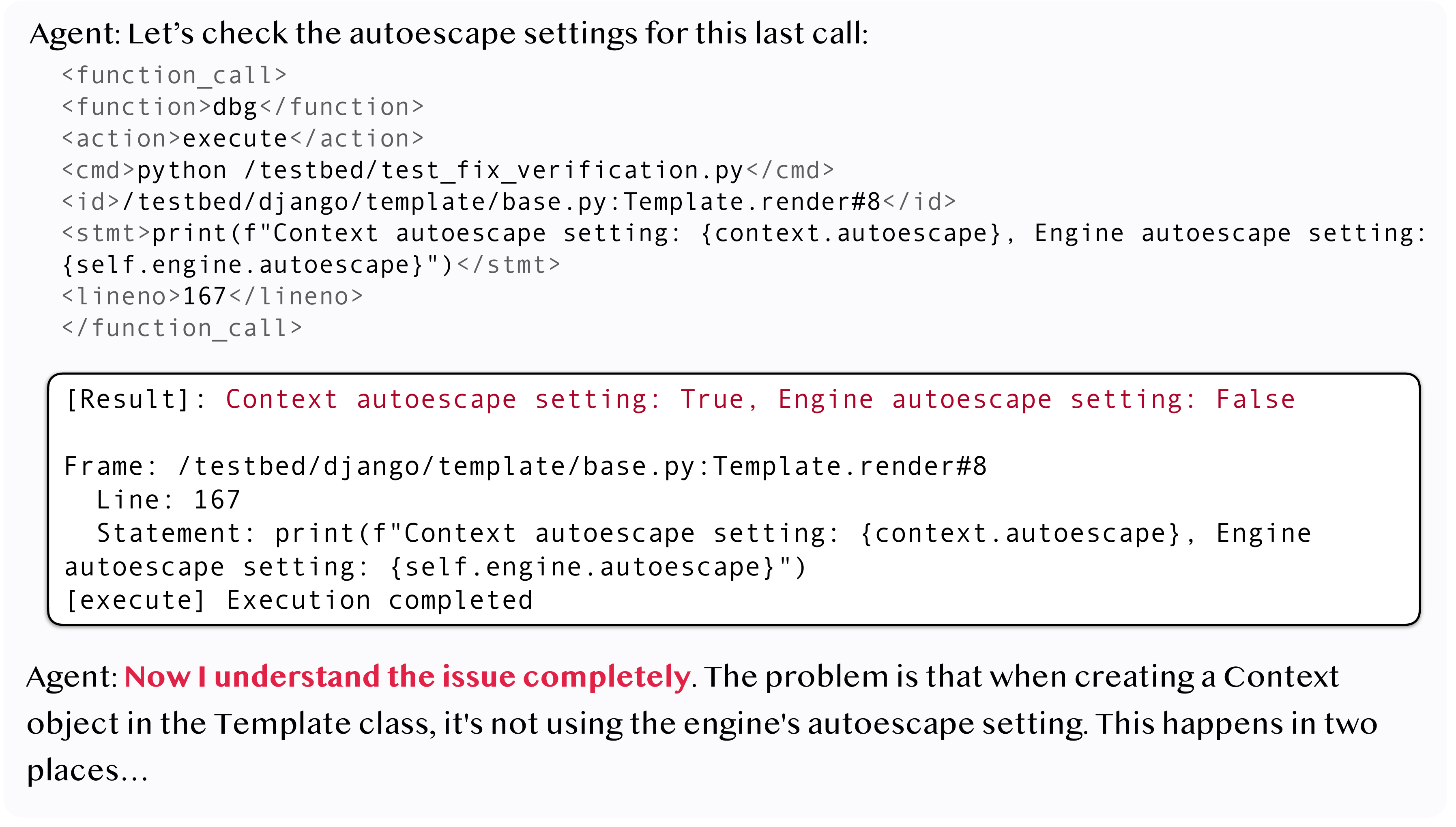}
    \caption{Diagnosing the state mismatch in \texttt{django-11119}~\cite{django11119}.}
    \label{fig:django-11119-casestudy}
\end{figure}

We present a case study on the \texttt{django-11119} task~\cite{django11119}, which involves a subtle, non-crashing bug where the template \texttt{Engine} fails to apply its \texttt{autoescape} setting, causing the template variables to be rendered without proper HTML escaping. 
This type of state-dependent issue is exceptionally difficult to diagnose with coarse-grained, post-mortem analysis, as the final output offers no clue about the internal state discrepancy. After navigating to the suspected function call, \dbgagent{} leverages the \texttt{execute} command to inject a diagnostic statement into a specific execution frame (\texttt{Template.render\#8}), allowing it to simultaneously inspect the \texttt{autoescape} attribute of both the local \texttt{Context} object and the parent \texttt{Engine} object.
As shown in Figure~\ref{fig:django-11119-casestudy}, this single, targeted query immediately revealed the critical state mismatch (\texttt{Context autoescape: True} vs. \texttt{Engine autoescape: False}), providing a precise fault localization that would be more costly and complex with traditional line-by-line debugging.

\mybox{Finding 3: The utilization of \adis{} scales with LLM capability, as stronger models invoke the interface more frequently. }

\input{tables/rq2-abaltion}

We further perform an ablation study on the components of our designed \flt{}. Specifically, we evaluate four settings on the \swebench{} Lite benchmark using \claudesonnet{}, starting from a baseline \baseagent{} and progressively adding \flt{} components ($\tau$, $args$, and $ret$) to form the full \dbgagent{}. As shown in Table~\ref{tab:flt-ablation}, each component contributes to the performance, with the number of resolved tasks increasing from 130 for the \baseagent{} to 152 for the \dbgagent{} under a moderate average cost increase from \$0.90 to \$1.26. 

%% file: tables/baseline.tex
{
\definecolor{table_shallow_blue}{RGB}{229, 238, 251}
\definecolor{table_white}{RGB}{255,255,255}

\begin{table*}[!h]
    \centering
    \caption{Performance of \dbgagent{} and  Baselines on \swebench{} Lite and Verified benchmarks.}
    \label{tab:rq1-sota}
    \begin{adjustbox}{width=0.97\textwidth}
    \begin{threeparttable}
    \setlength\tabcolsep{22pt}
    
        \begin{tabular}{ll|rr|r}

            \toprule

            \multirow{2}{*}{\textbf{Agent}}
                                                    & \multirow{2}{*}{\textbf{LLM}}
                                                    & \multicolumn{2}{c|}{\textbf{\% Resolved}}
                                                    & \multicolumn{1}{r}{\textbf{Avg.}}                                                                                                              \\
            \cmidrule(lr){3-4}
                                                    &                                           & \textbf{Verified}      & \multicolumn{1}{c|}{\textbf{Lite}} & \multicolumn{1}{r}{\textbf{\$ Cost}} \\

            \midrule

            \tools{}~\cite{anthropic2025claude3_7,toolclaude35}                    & \claudesonnet{}                         & 316\phantom{0}(63.2\%) & -              & 3.95\tnote{*}                       \\
                                        & \claudethreefive{}           & 245\phantom{0}(49.0\%) & -                                  & 3.11\tnote{*}                                \\

            \rowcolor{table_shallow_blue}
            \sweagent{}~\cite{sweagent,yang2025swe} & \claudesonnet{}                           & 291\phantom{0}(58.2\%) & 144\phantom{0}(48.0\%)             & -                                    \\
            \rowcolor{table_shallow_blue}
                                                    & \claudethreefive{}                        & 168\phantom{0}(33.6\%) & 69\phantom{0}(23.0\%)              & 1.62                                 \\
            \rowcolor{table_shallow_blue}
                                                    & \gpto{}                                   & 116\phantom{0}(23.2\%) & 55\phantom{0}(18.3\%)              & 2.53                                 \\

            \sweagentmini{}~\cite{swebench_website} & \claudesonnet{}                           & 264\phantom{0}(52.8\%) & -                                  & 1.43\tnote{*}        \\
                                                    & \gpto{}                                   & 108\phantom{0}(21.6\%) & -                                  & 0.98\tnote{*}                                  \\

            \rowcolor{table_shallow_blue}
            \agentless{}~\cite{agentless}           & \claudethreefive{}                        & 254\phantom{0}(50.8\%) & 122\phantom{0}(40.7\%)             & 1.19                                \\
            \rowcolor{table_shallow_blue}
                                                    & \gpto{}                                   & 194\phantom{0}(38.8\%) & 96\phantom{0}(32.0\%)              & 0.70                                 \\

            \acr{}~\cite{specrover}                 & \claudethreefive{}                        & 231\phantom{0}(46.2\%) & -                                  & 0.74                                    \\
                                                    & \gpto{}                                   & 192\phantom{0}(38.4\%) & 92\phantom{0}(30.7\%)              & 0.65                                    \\

            \rowcolor{table_shallow_blue}
            \ssi{}~\cite{swegpt}                    & \claudethreefive{}                        & 177\phantom{0}(35.4\%) & 71\phantom{0}(23.7\%)              & 0.42                                 \\
            \rowcolor{table_shallow_blue}
                                                    & \gpto{}                                   & 159\phantom{0}(31.8\%) & 62\phantom{0}(20.7\%)              & 0.78                                 \\
            \rowcolor{table_shallow_blue}
                                                    & Qwen2.5-instruct                          & 127\phantom{0}(25.4\%) & 54\phantom{0}(18.0\%)              & -                                    \\

            \midrule

            \textbf{\dbgagent{}}                    & \claudesonnet{}                           & 319\phantom{0}(63.8\%) & 152\phantom{0}(50.7\%)             & 1.28                                 \\
                                                    & \claudethreefive{}                        & 256\phantom{0}(51.2\%) & 111\phantom{0}(37.0\%)             & 0.58                                 \\
                                                    & \gpto{}                                   & 181\phantom{0}(36.2\%) & 82\phantom{0}(27.3\%)              & 0.93                                 \\
                                                    & \qwen{}                                   & 157\phantom{0}(31.4\%) & 63\phantom{0}(21.0\%)              & -                                    \\

            \bottomrule
        \end{tabular}
      
\begin{tablenotes}
  \footnotesize
  \item[]  "-" indicates data is not applicable or publicly available.  \  * indicates API cost is not publicly reported and is calculated from trajectories~\cite{swebench_website}.
\end{tablenotes}
    \end{threeparttable}
      \end{adjustbox}
\end{table*}

}

%% file: tables/rq1.tex
{
\definecolor{table_shallow_blue}{RGB}{229, 238, 251}
\definecolor{table_white}{RGB}{255,255,255}

\begin{table*}[!h]
    \centering
    \caption{Performance of \dbgagent{} and variants on \swebench{} Lite and Verified benchmarks.}
    \setlength\tabcolsep{10pt}
    \label{tab:rq1-ablation-result}
    \begin{adjustbox}{width=0.97\textwidth}
    \begin{threeparttable}
        \begin{tabular}{ll|cc|ccc|r}

            \toprule

            \multirow{2}{*}{\textbf{Agent}}
             & \multirow{2}{*}{\textbf{LLM}}
             & \multicolumn{2}{c|}{\textbf{\% Resolved}}
             & \multicolumn{3}{c|}{\textbf{\% Correct Location}}
             & \multicolumn{1}{r}{\textbf{Avg.}}                                                                                                                                 \\
            \cmidrule(lr){3-4} \cmidrule(lr){5-7}
             &                                                   & \textbf{Verified}                      & \textbf{Lite}
             & \multicolumn{1}{c}{\textbf{Line}}                & \multicolumn{1}{c}{\textbf{Function}} & \multicolumn{1}{c|}{\textbf{File}}
             & \multicolumn{1}{r}{\textbf{\$ Cost}}                                                                                                                              \\

            \midrule

            \rowcolor{table_shallow_blue}
            \cellcolor{table_white}
             & \claudesonnet{}                                   & 275\phantom{0}(55.0\%)                 & 130\phantom{0}(43.3\%)             & 47.0\% & 61.6\% & 77.0\% & 1.02 \\
             & \claudethreefive{}                                & 222\phantom{0}(44.4\%)                 & 92\phantom{0}(30.7\%)              & 44.4\% & 56.6\% & 73.6\% & 0.49 \\
            \rowcolor{table_shallow_blue}
            \cellcolor{table_white}
             & \gpto{}                                           & 163\phantom{0}(32.6\%)                 & 72\phantom{0}(24.0\%)              & 35.8\% & 49.6\% & 63.0\% & 0.87 \\
            \multirow{-4}{*}{\baseagent{}}
             & \qwen{}                                           & 146\phantom{0}(29.2\%)                 & 57\phantom{0}(19.0\%)              & 36.4\% & 46.0\% & 64.8\% & -    \\

            \midrule

            \rowcolor{table_shallow_blue}
            \cellcolor{table_white}
             & \claudesonnet{}                                   & 279\phantom{0}(55.8\%)                 & 133\phantom{0}(44.3\%)             & 46.4\% & 60.2\% & 76.4\% & 1.62 \\
             & \claudethreefive{}                                & 225\phantom{0}(45.0\%)                 & 96\phantom{0}(32.0\%)              & 44.8\% & 53.8\% & 72.0\% & 0.89 \\
            \rowcolor{table_shallow_blue}
            \cellcolor{table_white}
             & \gpto{}                                           & 156\phantom{0}(31.2\%)                 & 71\phantom{0}(23.7\%)              & 33.0\% & 43.6\% & 56.0\% & 1.09 \\
            \multirow{-4}{*}{\pdbagent{}}
             & \qwen{}                                           & 142\phantom{0}(28.4\%)                 & 60\phantom{0}(20.0\%)              & 38.0\% & 46.8\% & 63.2\% & -    \\

            \midrule
            \rowcolor{table_shallow_blue}
            \cellcolor{table_white}
             & \claudesonnet{}                                   & \textbf{319\phantom{0}(63.8\%)}        & \textbf{152\phantom{0}(50.7\%)}    & 51.2\% & 66.4\% & 78.8\% & 1.28 \\
             & \claudethreefive{}                                & \textbf{256\phantom{0}(51.2\%)}        & \textbf{111\phantom{0}(37.0\%)}    & 47.8\% & 58.6\% & 74.6\% & 0.58 \\
            \rowcolor{table_shallow_blue}
            \cellcolor{table_white}
             & \gpto{}                                           & \textbf{181\phantom{0}(36.2\%)}        & \textbf{82\phantom{0}(27.3\%)}     & 37.2\% & 51.2\% & 65.4\% & 0.93 \\
            \multirow{-4}{*}{\textbf{\dbgagent{}}}
             & \qwen{}                                           & \textbf{157\phantom{0}(31.4\%)}        & \textbf{63\phantom{0}(21.0\%)}     & 39.2\% & 49.0\% & 66.8\% & -    \\

            \bottomrule
        \end{tabular}
        \begin{tablenotes}
  \footnotesize
  \item[]  "-" indicates data is not applicable or publicly available.
\end{tablenotes}
        \end{threeparttable}
    \end{adjustbox}
\end{table*}

}

%% file: tables/rq3-adi_agent.tex
{
\definecolor{table_shallow_blue}{RGB}{229, 238, 251}
\definecolor{table_white}{RGB}{255,255,255}

\begin{table}[!tb]
\centering
\caption{Performance of SOTA Agents with and without \adis{} Integration on \swebench{} Verified benchmark}
\label{tab:sota_comparison}
\begin{adjustbox}{width=0.97\textwidth, center}
\begin{threeparttable}
    \setlength\tabcolsep{18pt}
    \begin{tabular}{ll r r r}
        \toprule
        \textbf{Approach} & \textbf{LLM} & \textbf{\#Resolved} & \textbf{Improv. (\%)} & \textbf{Avg. Cost (\$)} \\
        \midrule
        \sweagentmini{} & \claudesonnet{} & 264 & - &  1.43\\
        \sweagentmini{}$_{\text{\tiny ADI}}$ & \claudesonnet{} & 292 & 10.6 &  1.57\\
        \midrule
        \sweagentmini{} & \gpto{} & 108 & - & 0.98 \\
        \sweagentmini{}$_{\text{\tiny ADI}}$ & \gpto{} & 128 & 18.5 & 1.16 \\
        \midrule
        \acr{} & \claudethreefive{} & 231 & - & 0.74 \\
        \acr{}$_{\text{\tiny ADI}}$ & \claudethreefive{} & 248 & 7.3 & 0.86 \\
        \midrule
        \acr{} & \gpto{} & 192 & - &  0.65 \\
        \acr{}$_{\text{\tiny ADI}}$ & \gpto{} & 204 &  6.2  &  0.72 \\
        \bottomrule
    \end{tabular}
    \begin{tablenotes}
        \item[] \footnotesize Subscript "ADI" denotes the approach integrated with Agent-centric Debugging Interface.
    \end{tablenotes}
\end{threeparttable}
\end{adjustbox}
\end{table}

}

%% file: tables/rq2-abaltion.tex
\begin{table}[htb]
\caption{\flt{} component ablation on the \swebench{} Lite benchmark}
\label{tab:flt-ablation}
\centering
\setlength\tabcolsep{5pt}
\begin{adjustbox}{width=0.5\linewidth}
\begin{tabular}{lrr}
\toprule
\textbf{FLT component} & \textbf{\#Resolved} & \textbf{Avg.\$} \\
\midrule
Basic (\baseagent{}) & 130 & 0.90 \\
Basic+$\tau$ & 142 & 1.14 \\
Basic+$\tau$+$args$ & 145 & 1.23 \\
Basic+$\tau$+$args$+$ret$ (\dbgagent{}) & \textbf{152} & 1.26 \\
\bottomrule
\end{tabular}
\end{adjustbox}
\end{table}

%% file: threats.tex
\section{Threats to Validity}
\label{sec:threats}

\noindent \textbf{Threats to internal validity.}
We take several measures to ensure the internal validity of our \adi{}. First, we address the potential confounding effect of runtime overhead from our tracing mechanism. By using a selective instrumentation strategy, we limit tracing only to the function under inspection. Our measurements confirm this approach is effective, showing only a modest increase in average test execution time (0.68s to 0.87s) on the SWE-bench Verified set. 
\cmr{A potential threat to internal validity arises from \adi{}'s on-demand tracing design, which requires re-execution and may therefore introduce additional runtime overhead and sensitivity to non-deterministic executions. In practice, \adi{} re-executes the program 5.19 times per task on average, corresponding to about 4 seconds of total execution time per task on SWE-bench Verified. To mitigate this threat, we prompt the agent to generate minimal reproduction scripts as stable debugging entry points and restrict \adi{} from tracing external libraries; we did not observe noticeable non-deterministic effects in our evaluation.}
Then, to guarantee reproducibility and mitigate system-level variations, all evaluations are conducted using the official SWE-bench Docker environment, which encapsulates the exact repository state and dependencies for each task. To address the inherent stochasticity of LLMs, we set the model's temperature to 0 for all experiments. Furthermore, our evaluation is performed on large-scale benchmarks (\swebench{} Lite and Verified, with 300 and 500 tasks respectively), ensuring that our findings are stable. Any manual analysis, such as the case studies, was cross-validated by three authors to mitigate potential interpretation bias.

\noindent \textbf{Threats to external validity.}
The primary threat to external validity lies in the generalizability of our \adis{} and findings. We mitigate this by grounding our evaluation in the \swebench{} dataset, a widely-recognized benchmark composed of real-world software engineering tasks from popular open-source Python repositories. The authenticity of these tasks ensures our results reflect challenges faced in genuine development scenarios. Moreover, for our generality study (RQ2), we integrate \adis{} into two representative, state-of-the-art agents (\sweagentmini{} and \acr{}) chosen for their distinct architectures. This demonstrates that the benefits of \adis{} are not confined to our specific agent implementation but can extend to other existing frameworks.
\cmr{However, the generalizability of \adis{} may still vary across bug classes. In particular, \adis{} is currently better suited to bugs whose diagnosis depends on function-level data flow, control flow, and state mutations, while bugs involving concurrency effects, memory leaks, or interactions spanning many shallow calls may be less well supported by the current FLT abstraction.} 
\cmr{Moreover, as a modular debugging interface, \adis{} may also depend in part on how effectively a host agent invokes it. Our cross-agent evaluation on \sweagentmini{} and \acr{} partially mitigates this concern.}
\cmr{Additionally, although our current implementation is validated on Python, the core idea of \adis{} is not inherently Python-specific, as the stack frame is a fundamental abstraction across many programming languages and can be extended to other language ecosystems with appropriate instrumentation support.}

\noindent \textbf{Threats to construct validity.}
A potential threat to construct validity is whether our evaluation metrics accurately capture an agent's problem-solving capability. To address this, our primary metric is the Resolved Rate (Pass@1), the standard and most accepted metric for task completion in this domain. We complement this with two diagnostic metrics: the Correct Location Rate (\%) to assess the precision of the generated patches, and the Average Cost (\$) to measure the economic efficiency of the process. Collectively, these metrics provide a holistic assessment, evaluating not only the final correctness but also the accuracy and efficiency of the solution.

%% file: conclusion.tex
\section{Conclusion}
\label{sec:conclusion}

In this paper, we address the limitations of autonomous agents, which are hindered by coarse-grained execution feedback and the cost-inefficiency of traditional line-by-line debuggers. We introduce \adi{}, a novel agent-centric debugging interface that enables cost-efficient dynamic analysis through a function-level interaction model, powered by our \textit{\flt{}} and high-level navigational commands. Our evaluation on \swebench{} demonstrates that a basic agent equipped with \adi{} achieves \cmr{superior} performance, resolving 63.8\% of tasks on the \swebench{} Verified benchmark at a low average cost. Finally, we establish \adi{} as a general enhancement, showing that it delivers consistent performance gains when integrated as a plug-and-play component into the existing agents.

\section*{Data Availability}
The source code of our \adi{}, along with all agent implementations, evaluation scripts, and detailed results, are publicly available at our GitHub repository~\cite{githubrepo,xiang2026framepilot}.

\begin{acks}
This work is partially supported by the National Natural Science Foundation of China (Grant No. 62372220). It is also partially supported by Ant Group Research Fund.
\end{acks}